# Charting the Landscape of Oxygen Ion Conductors: A 60-Year Dataset with Interpretable Regression Models

*Seong-Hoon Jang*[*1], *Shin Kiyohara*[1], *Hitoshi Takamura*[2], *Yu Kumagai*[**1,3]

[1] *Institute for Materials Research*, Tohoku University, 2-1-1 Katahira, Aoba-ku, Sendai 980-8577, Japan.

[2] Department of Materials Science, Graduate School of Engineering, Tohoku University, 6-6-02 Aramaki, Aoba-ku, Sendai 980-8579, Japan.

[3] Organization for Advanced Studies, Tohoku University, 2-1-1 Katahira, Aoba-ku, Sendai 980-8577, Japan

*Corresponding authors:

jang.seonghoon.b4@tohoku.ac.jp (S.-H. Jang)

yukumagai@tohoku.ac.jp (Y. Kumagai)

ABSTRACT

Oxygen ion conductors are indispensable materials for such as solid oxide fuel cells, sensors, and membranes. Despite extensive research across diverse structural families, systematic data enabling comparative analysis remain scarce. Here, we present a curated dataset of oxygen ion conductors compiled from 84 experimental reports spanning 60 years, covering 483 materials. Each record includes activation energy ($E_a$) and prefactor ($A$) derived from Arrhenius plots, alongside detailed metadata on structure, composition, measurement method, and data source. When the original papers derive these using an erroneous Arrhenius equation $\sigma_T = A \exp\left(-\frac{E_a}{RT}\right)$, where ($\sigma_T$ is the oxygen ion conductivity at temperature $T$ and $R$ is the gas constant), we replotted these using the correct one, $\sigma_T T = A \exp\left(-\frac{E_a}{RT}\right)$. To illustrate how the database can be used, we constructed interpretable regression models for predicting oxygen ionic conductivity. Two symbolic regression models for $E_a$ and $A$ suggest that oxygen ion transport is primarily governed by local coordination environment and the electrostatic interactions, respectively. This dataset establishes a reliable foundation for data-driven discovery and predictive modeling of next-generation oxygen ion conductors.

**Background & Summary**

Oxygen ion conductors are key functional materials underpinning technologies such as solid oxide fuel cells, oxygen sensors, and catalytic membranes, where efficient ionic transport is essential for performance and sustainability.[1-8] Over the past six decades, a wide range of structural families, including perovskites/fluorites family,[9-45] Bi-based family (such as $\delta\text{-}Bi_2O_3$,[41, 46-54] BIMEVOX,[55-59] and $Bi_2UO_6$ classes[60]), complex framework family (such as scheelite,[61-72] apatite,[73-81] cuspidine,[82-84] melilite,[85-88] garnet,[89, 90] pyrochlore,[91] and columbite classes[92]) and $\alpha\text{-}SrSiO_3$ family,[93-96] have been reported to exhibit oxygen ion conductivity. Recent breakthroughs in $Ba_4Nb_4MoO_{20}$- and $\alpha\text{-}SrSiO_3$-based materials highlight the promise of structural diversity in enabling fast ion transport.[36, 96] However, the richness of these material families also poses a challenge: the structural and chemical space remains sparsely explored, and systematic design principles for new conductors are limited.

To address this gap, comprehensive data collection and curation are essential. Here, we present a manually curated dataset of oxygen ion conductors derived from 84 experimental reports, encompassing 483 unique oxides published over 60 years.[10-39, 41, 43-96] The dataset provides activation energies ($E_a$) and prefactors ($A$) extracted from Arrhenius-type analyses of ionic conductivity. When the original papers fit the conductivity using an erroneous Nernst-Einstein formula, we replot the conductivity and reevaluated these (see **Methods**). To examine the effectiveness of our database, we constructed interpretable two symbolic regression models that predict $E_a$ and $A$. These models enable the identification of dominant descriptors and offering physical insights into the factors governing oxygen ion transport. This dataset not only consolidates the historical knowledge base but also offers a foundation for data-driven discovery and design of next-generation oxygen ion conductors. In addition, our systematic data will be utilized for the validation of machine learning potentials, which have been developing rapidly in recent years.[97-99]

## Methods

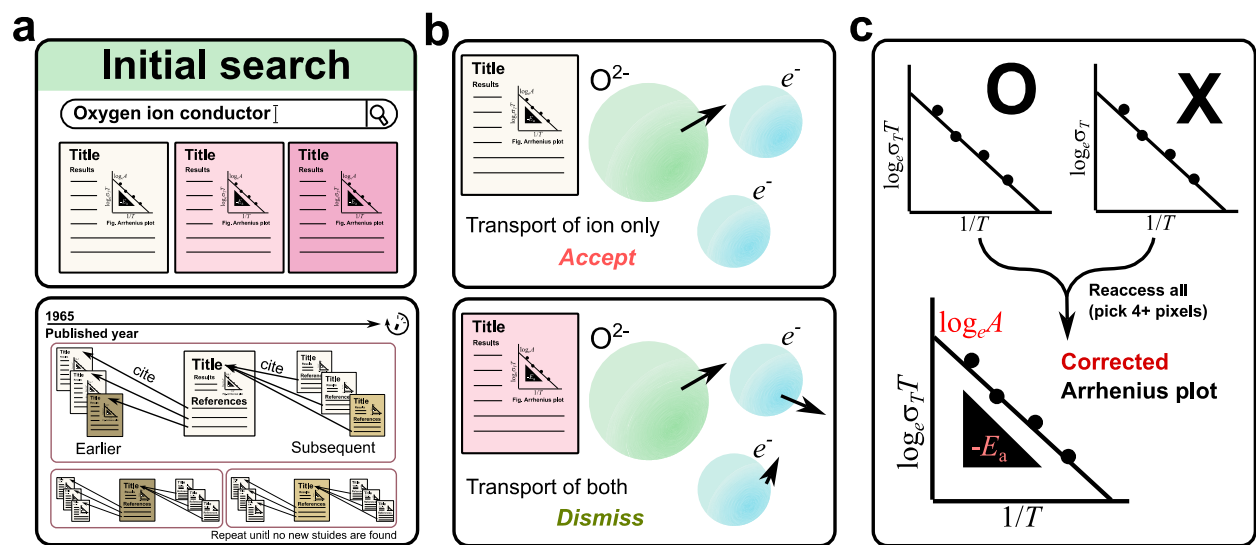

**Fig. 1** Workflow for dataset construction and curation of oxygen ion conductors. (a) Initial literature search using multiple academic databases identified candidate reports containing multi-temperature oxygen ion conductivities. Forward and backward citation searches were then performed iteratively until no additional relevant studies were found. (b) Conductivities dominated by electronic contributions were excluded to ensure that the dataset represents oxygen ion transport. (c) To correct for the widespread use of an incorrect Nernst-Einstein equation, all reported plots were manually re-examined (see text for details). For example, activation energies ($E_a$) and prefactors ($A$) were recalculated from at least four data points in figures to eliminate systematic biases and ensure consistency across the dataset.

To construct a comprehensive dataset of oxygen ion conductors, we systematically surveyed the literature spanning the past 60 years (**Fig. 1a**). We first used multiple academic search engines including Web of Science,[100] Scopus,[101] and Google Scholar[102] with keywords such as "oxygen ion conductivity", "Arrhenius plots", and "solid oxide fuel cells". From the identified reports containing oxygen ion conductivities at multiple temperatures ($\sigma_T$, where $T$ are measured temperatures), we then expanded our search by following two directions: (i) backward citations (references cited by these papers) and (ii) forward citations (subsequent papers citing them). This iterative process was repeated until no new experimental studies with relevant conductivity data were found. In this study, we targeted both purely oxygen ion conducting electrolytes and mixed ionic-electronic conductors; in the latter case, only the data in which the oxygen-ion conductivity was explicitly separated from the electronic contribution were included,[11, 12, 22, 31] and all others were excluded (**Fig. 1b**). Note that such mixed ionic–electronic conductors may include additional inaccuracies due to methodological assumptions/uncertainties in deriving oxygen-ion diffusivities.

A common issue in the literature is the use of an incorrect form of the Arrhenius plots. The correct form of Arrhenius equation is $\sigma_T T = A \exp\left(-\frac{E_a}{RT}\right)$, which is derived from $\sigma_T T = D_T \frac{cq}{k_B}$ and the Nernst-Einstein equation $D_T = A_D \exp\left(-\frac{E_a}{RT}\right)$, where $R$, $D_T$, $c$, $q$, $k_B$, and $A_D$ denote the gas constant, the self-diffusion coefficient, the density of charge carriers, the ionic charge, the Boltzmann constant, and the prefactor for $A_D$, respectively). However, many reports instead erroneously use $\sigma_T = A \exp\left(-\frac{E_a}{RT}\right)$. This inconsistency results in systematic errors in $E_a$ and $A$. To address this, all conductivity plots were carefully re-examined (**Fig. 1c**); an example is provided in the **Supplementary Information.** For figures, at least four points were manually extracted across the measured temperature range; for tables, a minimum of two data points were used. From these, $E_a$ and $A$ were recalculated to ensure consistency across the dataset. This labor-intensive manual curation is the most important contribution of this study, which allowed us to standardize parameters and eliminate inherited biases.

**Data Records**

The dataset comprises oxygen ion conductivity parameters extracted from 84 experimental reports spanning 60 years, covering 483 oxides. Each entry in the dataset corresponds to one oxide under a given measurement condition and includes $E_a$ and $A$ derived from the Arrhenius plots. For cases where the Nernst-Einstein relationship exhibits two distinct linear regions, both the low-temperature ($T < T^*$) and high-temperature ($T > T^*$) regimes are reported separately, with the break point temperature ($T^*$) explicitly recorded. To ensure consistency and reproducibility, each data entry is annotated with extensive metadata (**Table 1**).

**Table 1** Metadata fields for the curated dataset.

| Field | Description |
|---|---|
| References | Family name of the first author, journal title, volume, initial page number, and published year. |
| DOI | Digital Object Identifier of the source publication, enabling traceability. |
| Year | Publication year of the experimental report. |
| Class | Structural class of the material (e.g., perovskite, scheelite, apatite). |
| Formula | Reported chemical composition. |
| Parsed formula | Standardized composition represented by $A_s$–$B_s$(–$C_s$) site classification. For example, in the perovskite oxide $La_{0.95}Sr_{0.05}Ga_{0.95}Mg_{0.05}O_{3-\delta}$, La and Sr occupy the $A_s$-site, while Ga and Mg occupy the $B_s$-site of the perovskite lattice.[10] |
| $E_a$ / $A$ | Activation energy in meV and prefactor in $K \cdot S \cdot cm^{-1}$. When there are two distinct linear regions, values for the low-temperature regime are shown. |
| $E_{a,\text{HT}}$ / $A_{\text{HT}}$ | Activation energy and prefactor values for the high-temperature regime (if applicable). |
| $T^*$ | Transition temperature separating low- and high-temperature regimes (if applicable). |
| Type of $\sigma_T$ | Conductivity type, distinguishing bulk conductivity from total conductivity (bulk and grain boundary contributions). |
| Measurement | Experimental technique employed (e.g., two-probe AC, impedance spectroscopy). |
| Measurement temperature range | The range of temperatures over which experimental measurements were performed. |
| Source | Origin of the data within the publication (figure or table reference). |
| Plot type | Axis configuration used in the original Arrhenius representation (e.g., $\log_e \sigma_T$-$1000/T$) |

# Data Statics

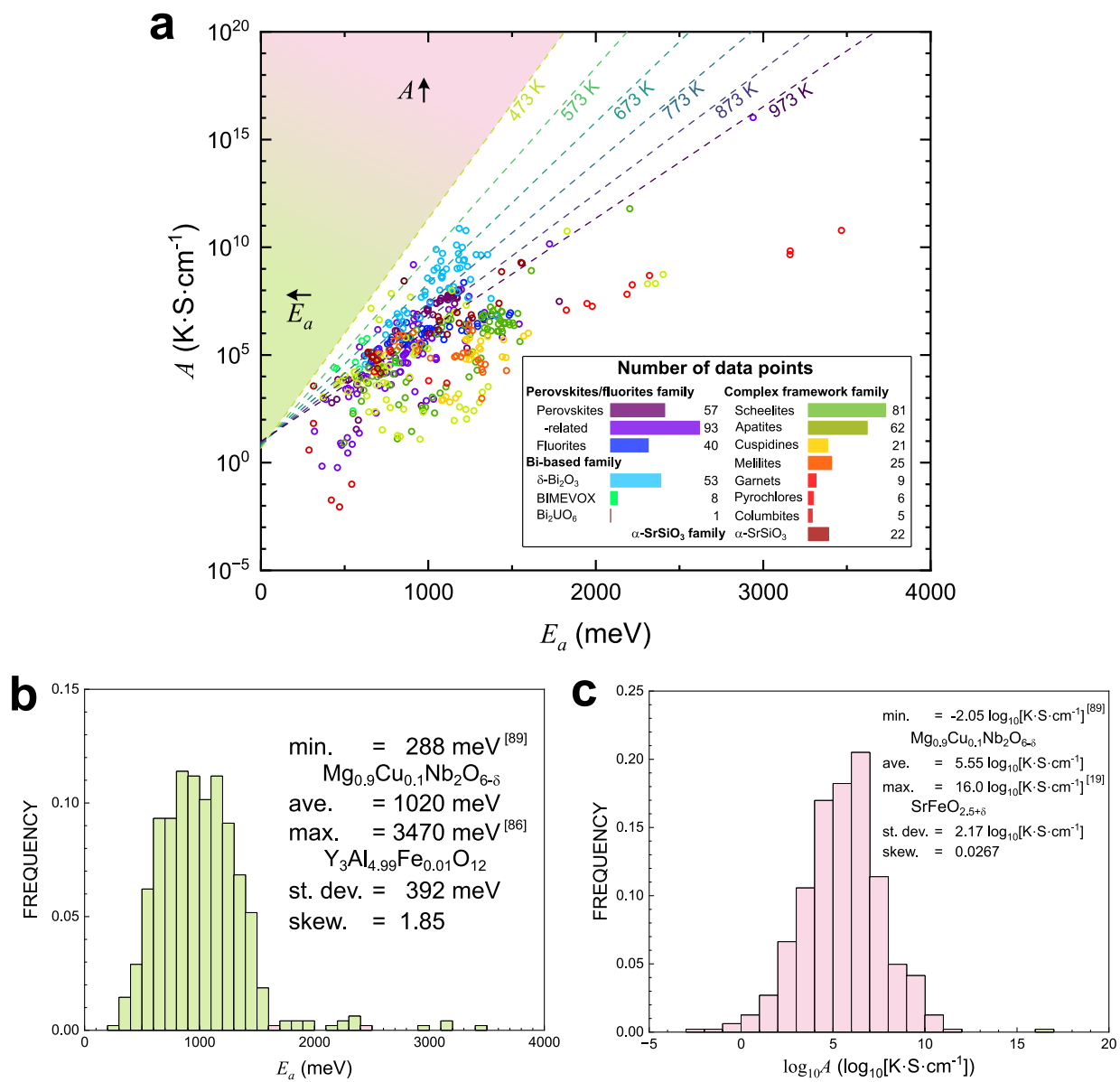

**Fig. 2** Overview of the curated dataset of oxygen ion conductivities. (a) Distribution of activation energy ($E_a$) and prefactor ($A$) for 483 oxides. We show $A$ in the low-temperature regime when two-region behavior was reported. Dashed lines indicate the condition required to achieve oxygen ionic conductivity $\sigma_T \geq 10^{-2}$ S·cm$^{-1}$ at different temperatures ($T = 473 - 873$ K). The sharded region ($T < 473$ K) highlights the particularly challenging region, which may be accessed either by suppressing $E_a$ (pink) or enhancing $A$ (green). The inset shows the classification of the dataset into 14 structural classes, along with the corresponding number of entries in each class. (b, c)
8

Histograms of (b) $E_a$ and (c) $\log_{10} A$. In (b) and (c), statistical parameters (minimum, average, maximum, standard deviation, and skewness) are provided; for minima and maxima, the experimental reports with their formula are referred to.[20, 89, 92]

**Fig. 2a** illustrates the distribution of data points across all oxides in the $A$-$E_a$ domain. For visualization, we used the low-temperature regime ($T < T^*$) when two-region behavior was reported. Dashed guidelines indicate the condition to achieve $\sigma_T \geq 10^{-2}$ S·cm$^{-1}$ at different temperatures. Notably, none of the investigated materials surpass this threshold at 473 K, underscoring the persistent challenge of achieving high ionic conductivity at intermediate temperatures. Access to this regime may be achieved either by reducing $E_a$ (pink-shaded region) or by increasing $A$ (green-shaded region). However, the weak positive correlation between $E_a$ and $A$ complicates this effort, underscoring the need to decouple descriptors for $E_a$ and $A$ in order to guide efficient materials design. Based on our dataset, the most effective oxygen ion conductors exhibiting $\sigma_T \geq 10^{-2}$ S·cm$^{-1}$ for $T = 473 - 573$ K are $\alpha$-SrSiO$_3$-type Sr$_{1.65}$Na$_{1.35}$Si$_3$O$_{8.325}$ ($E_a = 315$ meV and $A = 3.59 \times 10^3$ K·S·cm$^{-1}$),[95] apatite-type Nd$_{9.5}$Si$_{5.5}$Al$_{0.5}$O$_{26}$ ($E_a = 658$ meV and $A = 1.45 \times 10^7$ K·S·cm$^{-1}$) and Nd$_{9.83}$Si$_{4.5}$Al$_{1.5}$O$_{26}$ ($E_a = 779$ meV and $A = 7.05 \times 10^7$ K·S·cm$^{-1}$),[79] Bi$_2$UO$_6$ ($E_a = 856$ meV and $A = 2.69 \times 10^8$ K·S·cm$^{-1}$),[60] and BIMEVOX-type Bi$_2$V$_{0.9}$Mg$_{0.1}$O$_{5.35}$ ($E_a = 911$ meV and $A = 1.54 \times 10^9$ K·S·cm$^{-1}$).[93]

The oxides in the dataset are classified into 14 structural classes, namely perovskite, perovskite-related, fluorite, $\delta$-Bi$_2$O$_3$, BIMEVOX, Bi$_2$UO$_6$, scheelite, apatite, cuspidine, melilite, garnet, pyrochlore, columbite, and $\alpha$-SrSiO$_3$ classes. The counts of oxides per class are shown in the inset of **Fig. 2a**. Perovskites and related compounds constitute the largest fraction of reported data, reflecting the dominant research focus in the field, followed by scheelites, apatites, and other structures.

**Fig. 2b** shows the distribution of $E_a$, with most values lying between 200 and 2000 meV and an average of 1020 meV. The lowest $E_a$ (288 meV) was observed for the columbite Mg$_{0.9}$Cu$_{0.1}$Nb$_2$O$_{6-\delta}$,[92] while the highest (3470 meV) was reported for the garnet Y$_3$Al$_{4.99}$Fe$_{0.01}$O$_{12}$.[89] **Fig. 2c** presents the distribution of $\log_{10} A$, with most $A$ values lying between 1 and $10^{10}$ K·S·cm$^{-1}$ and an average of $10^{5.55} = 3.55 \times 10^5$ K·S·cm$^{-1}$. The lowest $A$ ($10^{-2.05} = 8.91 \times 10^{-3}$ K·S·cm$^{-1}$) was observed for the columbite $Mg_{0.8}Li_{0.2}Nb_2O_{6-\delta}$,[92] while the highest ($10^{16.0}$ K·S·cm$^{-1}$) was reported for the perovskite-related brownmillerite $SrFeO_{2.5+\delta}$.[20]

We also evaluated the reproducibility of the conductivities. The conductivities of 26 oxides were reported in more than one publication, providing opportunities for cross-validation. The average standard deviations were 78.1 meV for $E_a$ and $10^{0.41} = 2.58$ K·S·cm$^{-1}$ for $A$. Considering the distributions of $A$ and $E_a$ shown in **Figs. 2b** and **2c**, respectively, a high degree of consistency among independent studies is achieved despite variations in synthesis and measurement conditions.

**Technical Validation**

To support the reliability and scientific utility of the dataset, we performed symbolic regression analyses on both $E_a$ and $A$ values by using *GoodRegressor*, developed to derive analytical models that link experimental datasets with the structural and compositional features of materials. A brief summary of the program package is included in the **Supplementary Information,** whereas the technical and numerical details are provided in Ref. 103.

**Table 2** Structural, chemical, and physical properties of constituent elements of oxides for regression models, given as "features" for symbolic regression modeling.

| Properties | Description | Unit |
|---|---|---|
| $\bar{O}$ | Molar ratio of oxygen ions to metal ions | - |
| $M$ | Atomic mass | $g \cdot mol^{-1}$ |
| $Z$ | Valence | - |
| $r_{VI}$ | Shannon ionic radius with sixfold coordination to oxygen[104, 105] | Å |
| $r$ | Shannon ionic radius depending on $n_c$ (see below)[104, 105] | Å |
| $B$ | Bulk modulus | GPa |
| $G$ | Shear modulus | GPa |
| $\rho$ | Density | $g \cdot cm^{-3}$ |
| $\rho_{mol}$ | Molar density, defined as $\rho/M$ | $mol \cdot cm^{-3}$ |
| $\eta_f$ | Ionic filling rate (per unit volume), defined as $\eta_f = N_{avo}\rho_{mol}\left(\frac{4}{3}\pi r_{VI}^3\right)$ where $N_{avo}$ is the Avogadro's constant | - |
| $n$ | Principal quantum number of valence electrons | - |
| $l$ | Azimuthal quantum number of valence electrons | - |
| $\alpha$ | Thermal expansion coefficient (linear not volumetric) | $K^{-1}$ |

| $\kappa$ | Thermal conductivity | $W \cdot m^{-1} \cdot K^{-1}$ |
|---|---|---|
| $\chi - \chi_O$ | Difference in electronegativity between metal ions and oxygen | - |
| $\nu$ | Poisson's ratio | - |
| $\theta_D$ | Debye temperature | K |
| $n_c$ | Coordination number to oxygen ions, which depends on the occupied sites ($A_s$, $B_s$, or $C_s$) in different crystal classes. For example, in perovskite oxides, $n_c$ for $A_s$ and $B_s$ are given as 12 and 6, respectively. | - |
| $\langle \cdots \rangle$ | Average over the constituent metal ions | Common to $\cdots$ |
| $\sigma(\cdots)$ | Standard deviation over the constituent metal ions | Common to $\cdots$ |
| $r(\cdots)$ | Skewness over the constituent metal ions | - |

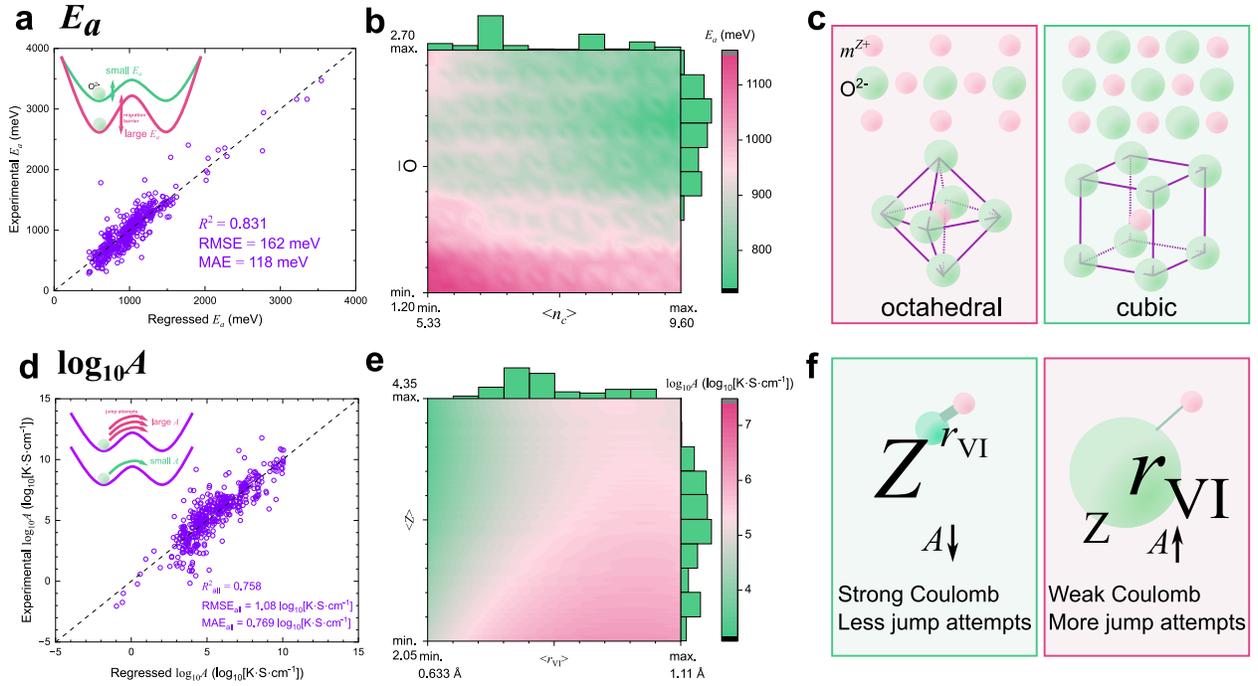

**Fig. 3** Symbolic regression models governing factors of oxygen ion transport. (a) Experimental versus regressed values for the symbolic regression model for $E_a$, with the schematic illustrations of potential energy surfaces exhibiting low (green curve) and high $E_a$ (pink curve). (b) Two-way partial dependence plot of the most important interaction for $E_a$: $\langle n_c \rangle$-$\bar{O}$ (see text for details). (c) Illustration of high (pink panel) and low (green panel) $E_a$, which depends on $\langle n_c \rangle$ and $\bar{O}$ both. (d) Experimental versus regressed values for the symbolic regression model for $\log_{10} A$, with the schematic illustrations of potential energy surfaces with low (green arrows) and high $\log_{10} A$ (pink arrow). (e) Two-way partial dependence plot of the most important interaction for $\log_{10} A$: $\langle r_{VI} \rangle$-$\langle Z \rangle$. (f) Illustration of how $\langle r_{VI} \rangle$ and $\langle Z \rangle$ modulate Coulombic interactions and oxygen ion jump attempts. In (b) and (e), histograms of the corresponding features for the full dataset are also shown along the upper and right axes.

As shown in **Fig. 3a**, the symbolic regression model for $E_a$ achieved $R^2 = 0.831$, RMSE = 162 meV, and MAE = 118 meV across the full dataset. The key interactions distinguishing oxides with suppressed $E_a$ (green curve) from those with large $E_a$ (pink curve) were identified (see Ref. 103 for details). The most important interaction was found between the average metal-oxygen coordination number ($\langle n_c \rangle$) and the molar ratio of oxygen to metal ions ($\bar{O}$). Note that the most important interaction does not necessarily exclude the contribution of the other features. The two-way partial dependence plot in **Fig. 3b**, obtained by varying $\langle n_c \rangle$ and $\bar{O}$ from the minimum to the maximum values of the full dataset while fixing other features at their mean values. As illustrated in **Fig. 3c,** a clear trend emerges: oxygen-rich environments with higher coordination numbers $\langle n_c \rangle$ (such as cubic coordination) become more favorable than oxygen-poor environments with lower $\langle n_c \rangle$ (such as octahedral or tetrahedral coordination), probably because the increased O-O repulsion in high-$\langle n_c \rangle$ configurations expands and softens the local oxygen sublattice, thereby facilitating oxygen-ion migration.

As shown in **Fig. 3d**, the symbolic regression model for $\log_{10} A$ achieved $R^2 = 0.758$, RMSE $= 10^{1.07}$ K·S·cm$^{-1}$, and MAE $= 10^{0.769}$ K·S·cm$^{-1}$ for the entire dataset. The principal interactions distinguishing low and high $A$ values (green and pink arrows, respectively) were identified (see also Ref. 103 for details). The most important interaction involves the average Shannon ionic radius with sixfold coordination ($\langle r_{VI} \rangle$)[104, 105] and the average metal ion valence ($\langle Z \rangle$), whose two-way dependence is shown in **Fig. 3e**. A straightforward interpretation follows; small $\langle r_{VI} \rangle$ combined with large $\langle Z \rangle$ strengthens the metal-oxygen Coulombic attraction, which stabilizes oxygen ions at their lattice sites and reduces their configurational freedom. As a result, the rate of attempted ionic jumps is suppressed. Conversely, when the metal-oxygen attraction is weaker, oxygen ions experience less localization and can attempt jumps more frequently, leading to the higher $\log_{10} A$ (**Fig. 3f**).

Our regression models can also be utilized for materials design. By substituting or doping ions and adjusting compositions, the model enables the exploration of new compounds with desirable target metrics, characterized by reduced $E_a$ and enhanced $A$. In this study, the most promising candidate was identified as the apatite-type compound $La_{9.5}Si_{5.5}Al_{0.5}O_{26}$. The predicted $E_a$ and $A$ were 494 meV and $7.41 \times 10^6$ K·S·cm$^{-1}$, respectively. This composition is a modified analogue of the experimentally reported apatite $Nd_{9.5}Si_{5.5}Al_{0.5}O_{26}$, which exhibited $E_a = 658$ meV and $A = 1.44 \times 10^7$ K·S·cm$^{-1}$.[79] This example demonstrates the utility of the curated dataset and regression framework as a quantitative guide for both experimental and theoretical materials discovery. We envision that such data-driven predictions will accelerate the development of next-generation oxygen-ion conductors.

**Associated content**

Supplementary Information: Correction of the Nernst-Einstein Relation in Arrhenius Plot Analysis and A Brief Summary of Symbolic Regression Tool *GoodRegressor*

**Author information**

**Data Availability**

The total dataset created in this study is openly available at https://github.com/JerryGarcia1995/OxygenIonConductor

**Code Availability**

The source code supporting materials prediction and design in this study is openly available at https://github.com/JerryGarcia1995/OxygenIonConductor

**Acknowledgments**

# References

1. Singh, M., Zappa, D. & Comini, E. Solid oxide fuel cell: Decade of progress, future perspectives and challenges. *Int. J. Hydrog. Energy* **46,** 27643–27674 (2021).

2. Li, J. *et al*. Advancements in solid oxide fuel cell technology: Bridging performance gaps for enhanced environmental sustainability. *Adv. Energ. Sust. Res.* **5,** 2400132 (2024).

3. Bolvin, J. C. & Mairesse, G. Recent materials developments in fast oxide ion conductors. *Chem. Mater.* **10,** 2870–2888 (1998).

4. Kharton, V., Marques, F. & Atkinson, A. Transport properties of solid oxide electrolyte ceramics: A brief review. *Solid State Ion.* **174,** 135–149 (2004).

5. Amow, G. & Skinner, S. J. Recent developments in Ruddlesden–Popper nickelate systems for solid oxide fuel cell cathodes. *J. Solid State Electrochem.* **10,** 538–546 (2006).

6. Shen, M., Ai, F., Ma, H., Xu, H. & Zhang, Y. Progress and prospects of reversible solid oxide fuel cell materials. *iScience* **24,** 103464 (2021).

7. Yang, X., Fernández-Carrión, A. J. & Kuang, X. Oxide ion-conducting materials containing tetrahedral moieties: Structures and conduction mechanisms. *Chem. Rev.* **123,** 9356–9396 (2023).

8. Zhang, W. & Yashima, M. Recent developments in oxide ion conductors: Focusing on Dion–Jacobson phases. *Chem. Commun.* **59,** 134–152 (2023).

9. Ishihara, T., Matsuda, H. & Takita, Y. Doped LaGaO3 perovskite type oxide as a new oxide ionic conductor. *J. Am. Chem. Soc.* **116,** 3801–3803 (1994).

10. Huang, K., Tichy, R. S. & Goodenough, J. B. Superior perovskite oxide-ion conductor; strontium- and magnesium-doped $LaGaO_3$: I, phase relationships and electrical properties. *J. Am. Ceram. Soc.* **81,** 2565–2575 (1998).

11. Ullmann, H., Trofimenko, N., Tietz, F., Stöver, D. & Ahamad-Khanlou. A. Correlation between thermal expansion and oxide ion transport in mixed conducting perovskite-type oxides for SOFC cathodes. *Solid State Ion.* **138,** 79–90 (2000).

12. Bucher, E., Egger, A., Ried, P., Sitte, W. & Holtappels, P. Oxygen nonstoichiometry and exchange kinetics of $Ba_{0.5}Sr_{0.5}Co_{0.8}Fe_{0.2}O_{3-\delta}$. *Solid State Ion.* **179,** 21–26 (2008).

13. Li, M. *et al.* A family of oxide ion conductors based on the ferroelectric perovskite $Na_{0.5}Bi_{0.5}TiO_3$. *Nat. Mater.* **13,** 31–35 (2014).

14. Yang, F., Zhang, H., Li, L., Reaney, I. M. & Sinclair, D. C. High ionic conductivity with low degradation in A-site strontium-doped nonstoichiometric sodium bismuth titanate perovskite. *Chem. Mater.* **28**, 5269–5273 (2016).

15. Goodenough, J. B., Ruiz-Diaz, J. E. & Zhen, Y. S. Oxide-ion conduction in $Ba_2InO_5$ and $Ba_3In_2MO_8$ (M=Ce, Hf, or Zr). *Solid State Ion.* **44,** 21–31 (1990).

16. Kuramochi, H., Mori, T., Yamamura, H., Kobayashi, H. & Mitamura, T. Preparation and conductivity of $Ba_2In_2O_5$ ceramics. *J. Ceram. Soc. Jpn.* **102,** 1159–1162 (1994).

17. Kurek, P., Bogusz, W., Jakubowski, W. & Krok, F. Impedance study of BIMGVOX ceramics. *Ionics* **2,** 474–477 (1996).

18. Uchimoto, Y., Yao, T., Takahi, H., Inagaki, T. & Yoshida, H. Crystal structure of $(Ba_{1-x}La_x)_2In_2O_{5+x}$ and its oxide ion conductivity. *Electrochemistry* **68,** 531–533 (2000).

19. Yao, T., Uchimoto, Y., Kinuhata, M., Inagaki, T. & Yoshida, H. Crystal structure of Ga-doped $Ba_2In_2O_5$ and its oxide ion conductivity. *Solid State Ion.* **132,** 189–198 (2000).

20. Patrakeev, M. V., Leonidov, I. A., Kozhevnikov, V. L. & Kharton, V. V. Ion–electron transport in strontium ferrites: relationships with structural features and stability. *Solid State Sci.* **6,** 907–913 (2004).

21. Shin, J. F., Oera, A., Apperley, D. C. & Slater, P. R. Oxyanion doping strategies to enhance the ionic conductivity in $Ba_2In_2O_5$. *J. Mater. Chem.* **21,** 874–879 (2011).

22. Ishihara, T. Oxide ion conductivity in defect perovskite, $Pr_2NiO_4$ and its application for solid oxide fuel cells. *J. Ceram. Soc. Jpn.* **122,** 179–186 (2014).

23. Fujii, K. *et al*. New perovskite-related structure family of oxide-ion conducting materials $NdBaInO_4$. *Chem. Mater.* **26,** 2488–2491 (2014).

24. López, C. A., Pedregosa, J. C., Lama, D. G. & Alonso, J. A. The strongly defective double perovskite $Sr_{11}Mo_4O_{23}$: Crystal structure in relation to ionic conductivity. *J. Appl. Cryst.* **47,** 1395–1401 (2014).

25. Fujii, K. *et al*. Improved oxide-ion conductivity of $NaBaInO_4$ by Sr doping. *J. Mater. Chem. A* **3,** 11985–11990 (2015).

26. Fop, S. *et al.* Oxide ion conductivity in the hexagonal perovskite derivative $Ba_3MoNbO_{8.5}$. *J. Am. Chem. Soc.* **138,** 16764–16769 (2016).

27. Yang, X., Liu, S., Lu, F., Xu, J. & Kuang, X. Acceptor doping and oxygen vacancy migration in layered perovskite $NdBaInO_4$-based mixed conductors. *J. Phys. Chem. C* **120,** 6416–6426 (2016).

28. Shiraiwa, M. Crystal structure and oxide-ion conductivity of $Ba_{1+x}Nd_{1-x}InO_{4-x/2}$. *J. Electrochem. Soc.* **164,** F1392–F1399 (2017).

29. McCombie K. S. *et al.* The crystal structure and electrical properties of the oxide ion conductor $Ba_3WNbO_{8.5}$. *J. Mater. Chem. A* **6,** 5290–5295 (2018).

30. Long, C. *et al*. High oxide ion conductivity in layer-structured $Bi_4Ti_3O_{12}$-based ferroelectric ceramics. *J. Mater. Chem. C* **7,** 8825–8835 (2019).

31. Song, J., Ning, D., Boukamp, B., Bassat, J.-M. & Bouwmeester, H. J. M. Structure, electrical conductivity and oxygen transport properties of Ruddlesden–Popper phases $Ln_{n+1}Ni_nO_{3n+1}$ (Ln = La, Pr and Nd; $n$ = 1, 2 and 3). *J. Mater. Chem. A* **8,** 22206–22221 (2020).

32. Fop, S. *et al.* High oxide ion and proton conductivity in a disordered hexagonal perovskite. *Nat. Mater.* **19,** 752–757 (2020).

33. Zhang, W. *et al*. Oxide-ion conduction in the Dion–Jacobson phase $CsBi_2Ti_2NbO_{10-\delta}$. *Nat. Commun.* **11,** 1224 (2020).

34. Fop, S., McCombie, K., Smith, R. I. & Mclaughlin A. C. Enhanced oxygen ion conductivity and mechanistic understanding in $Ba_3Nb_{1-x}V_xMoO_{8.5}$. *Chem. Mater.* **32,** 4724–4733 (2020).

35. Gilane, A. Fop, S., Sher, F., Smith, R. I. & Mclaughlin, A. C. The relationship between oxide-ion conductivity and cation vacancy order in the hybrid hexagonal perovskite $Ba_3VWO_{8.5}$. *J. Mater. Chem. A* **8,** 16506–16514 (2020).

36. Yashima, M. *et al*. High oxide-ion conductivity through the interstitial oxygen site in $Ba_7Nb_4MoO_{20}$-based hexagonal perovskite related oxides. *Nat. Commun.* **12,** 556 (2021).

37. Sakuda, Y., Hester, J. R. & Yashima, M. Improved oxide-ion and lower proton conduction of hexagonal perovskite-related oxides based on $Ba_7Nb_4MoO_{20}$ by $Cr^{6+}$ doping. *J. Ceram. Soc. Jpn.* **130,** 442–447 (2022).

38. Murakami, T. *et al*. High oxide-ion conductivity in a hexagonal perovskite-related oxide $Ba_7Ta_{3.7}Mo_{1.3}O_{20.15}$ with cation site preference and interstitial oxide ions. *Small* **18,** 2106785 (2022).

39. Strickler, D. W. & Carlson, W. G. Electrical conductivity in the $ZrO_2$-rich region of several $M_2O_3$–$ZrO_2$ systems. *J. Am. Ceram. Soc.* **48,** 286–289 (1965).

40. Tuller, H. L. & Nowick, A. S. Doped ceria as a solid oxide electrolyte. *J. Electrochem. Soc.* **122,** 255–259 (1975).

41. Miyayama, M., Nishi, T. & Yanagida, H. Oxygen ionic conduction in $Y_2O_3$-stabilized $Bi_2O_3$ and $ZrO_2$ composites. *J. Mater. Sci.* **22,** 2624–2628 (1987).

42. Yahiro, H., Eguchi, Y., Eguchi, K. & Arai, H. Oxygen ion conductivity of the ceria-samarium oxide system with fluorite structure. *J. Appl. Electrochem.* **18,** 527–531 (1988).

43. Mori, M. *et al*. Cubic-stabilized zirconia and alumina composites as electrolytes in planar type solid oxide fuel cells. *Solid State Ion.* **74,** 157–164 (1994).

44. Steele, B. C. H. Appraisal of $Ce_{1-y}Gd_yO_{2-y/2}$ electrolytes for IT-SOFC operation at 500°C. *Solid State Ion.* **129,** 95–110 (2000).

45. Kharton V. V. *et al*. Ceria-based materials for solid oxide fuel cells. *J. Mater. Sci.* **36,** 1105–1117 (2001).

46. Takahashi, T., Iwahara, H. & Arao, T. High oxide ion conduction in sintered oxides of the system $Bi_2O_3$-$Y_2O_3$. *J. Appl. Electrochem.* **5,** 187–195 (1975).

47. Takahashi, T. & Iawahara, H. Oxide ion conductors based on bismuthsesquioxide. *Mat. Res. Bull.* **13,** 1447–1453 (1978).

48. Verkerk, M. J., Keizer, K. & Burggraaf, A. J. High oxygen ion conduction in sintered oxides of the $Bi_2O_3$-$Er_2O_3$ system. *J. Appl. Electrochem.* **10,** 81–90 (1980).

49. Verkerk, M. J. & Burggraaf, A. J. High oxygen ion conduction in sintered oxides of the $Bi_2O_3$-$Ln_2O_3$ system. *Solid State Ion.* **3–4,** 463–467 (1981).

50. Benkaddour, M., Obbade, S., Conflant, P. & Drache, M. $Bi_{0.85}Ln_{0.15(1-n)}V_{0.15n}O_{1.5+0.15n}$ fluorite type oxide conductors: Stability, conductivity, and powder crystal structure investigations. *J. Solid State Chem.* **163,** 300–307 (2002).

51. Punn, R., Feteira, A. M., Sinclair, D. C. & Greaves, C. Enhanced oxide ion conductivity in stabilized $\delta$-$Bi_2O_3$. *J. Am. Chem. Soc.* **128,** 15386–15387 (2006).

52. Jung, D. W., Duncan, K. L. & Wachsman, E. D. Effect of total dopant concentration and dopant ratio on conductivity of $(DyO_{1.5})_x$–$(WO_3)_y$–$(BiO_{1.5})_{1-x-y}$. *Acta Mater.* **58,** 355–363 (2010).

53. Kunag, X., Payne, J. L., Johnson, M. R. & Evans, I. R. Remarkably high oxide ion conductivity at low temperature in an ordered fluorite-type superstructure. *Angew. Chem.* **51,** 690–694 (2012).

54. Karmalkar, D. N. *et al*. Enhanced phase stability and oxide-ion conductivity in V- and Sr/Ca-codoped $Bi_2O_3$ ceramics. *J. Phys. Chem. C* **129,** 107–120 (2024).

55. Abraham, F., Boivin, J. C., Mairesse, G. & Nowogrocki, G. The bimevox series: A new family of high performances oxide ion conductors. *Solid State Ion.* **40–41,** 934–937 (1990).

56. Sharma, V., Shukla, A. K. & Gopalakrishnan, J. Effect of aliovalent-cation substitution on the oxygen-ion conductivity of $Bi_4V_2O_{11}$. *Solid State Ion.* **58,** 359–362 (1992).

57. Yan, J. & Greenblatt, M. Ion conductivities of $Bi_4V_{2-x}M_xO_{11-x2}$ (M=Ti, Zr, Sn, Pb) solid solutions. *Solid State Ion.* **81,** 225–233 (1995).

58. Kim, S.-K. & Miyayama, M. Anisotropy in oxide ion conductivity of $Bi_4V_{2-x}Co_xO_{11-\delta}$. *Solid State Ion.* **104,** 295–302 (1997).

59. Yaremchenko, A. A. *et al*. Structure and electronic conductivity of $Bi_{2-x}La_xV_{0.9}Cu_{0.1}O_{5.5-\delta}$. *Mater. Chem. Phys.* **77,** 552–558 (2003).

60. Bonanos, N. High oxide ion conductivity in bismuth uranate, $Bi_2UO_6$. *Mat. Res. Bull.* **24,** 1531–1540 (1989).

61. Esaka, T., Mina-ai, T. & Iwahara, H. Oxide ion conduction in the solid solution based on the scheelite-type oxide $PbWO_4$. *Solid State Ion.* **57,** 319–325 (1992).

62. Esaka, T., Tachibana, R. & Takai, S. Oxide ion conduction in the Sm-substituted $PbWO_4$ phases. *Solid State Ion.* **92,** 129–133 (1996).

63. Cheng, J., Liu, C., Cao, W., Qi, M. & Shao, G. Synthesis and electrical properties of scheelite $Ca_{1-x}Sm_xMoO_{4+\delta}$ solid electrolyte ceramics. *Mat. Res. Bull.* **46,** 185–189 (2011).

64. Li, C., Bayliss, R. D. & Skinner, S. J. Crystal structure and potential interstitial oxide ion conductivity of $LnNbO_4$ and $LnNb_{0.92}W_{0.08}O_{4.04}$ (Ln = La, Pr, Nd). *Solid State Ion.* **262,** 530–535 (2014).

65. Takai, S. *et al*. Electrochemical properties of Cs-substituted $CaWO_4$ and $BaWO_4$ oxide ion conductors. *J. Ceram. Soc. Japan* **124,** 819–822 (2016).

66. Cheng, J. & He, J. Electrical properties of scheelite structure ceramic electrolytes for solid oxide fuel cells. *Mater. Lett.* **209,** 525–527 (2017).

67. Yang, X. *et al*. Cooperative mechanisms of oxygen vacancy stabilization and migration in the isolated tetrahedral anion Scheelite structure. *Nat. Commun.* **9,** 4484 (2018).

68. Kawaguchi, R., Akizawa, R., Shan, Y. J., Tezuka, K. & Katsumata, T. Synthesis and examination of $GdNb_{1-x}W_xO_{4+\delta}$ new scheelite-type oxide-ion conductor. *Solid State Ion.* **355,** 115415 (2020).

69. Auckett, J. E., Lopez-Odriozola, L., Clark, S. J. & Evans, I. R. Exploring the nature of the fergusonite–scheelite phase transition and ionic conductivity enhancement by $Mo^{6+}$ doping in $LaNbO_4$. *J. Mater. Chem. A* **9,** 4091–4102 (2021).

70. Yang, X. *et al*. Oxide-ion conductivity optimization in $BiVO_4$ scheelite by an acceptor doping strategy. *Inorg. Chem. Front*. **9,** 2644–2658 (2022).

71. Shan, Y. J., Kawaguchi, R., Akizawa, R. & Tezuka, K. Crystal structure and ionic conductivity of novel rare-earth niobates $LnNbO_4$ ($Ln$ = Nd, Sm, Eu, Gd) by substituting Nb with W. *Ionics* **29,** 2697–2703 (2023).

72. Mullens, B. G. *et al*. Variable temperature in situ neutron powder diffraction and conductivity studies of undoped $HoNbO_4$ and $HoTaO_4$. *Chem. Mater.* **36,** 5002–5016 (2024).

73. Nakayama S., Kageyama, T., Aono, H. & Sadaoka, Y. Ionic conductivity of lanthanoid silicates, $Ln_{10}(SiO_4)_6O_3$ (Ln = La, Nd, Sm, Gd, Dy, Y, Ho, Er and Yb). *J. Mater. Chem.* **5,** 1801–1805 (1995).

74. Abram, E. J., Sinclair, D. C. & West, A. R. A novel enhancement of ionic conductivity in the cation-deficient apatite $La_{9.33}(SiO_4)_6O_2$. *J. Mater. Chem.* **11,** 1978–1979 (2001).

75. Shaula, A. L., Kharton, V. V. & Marques, F. M. B. Oxygen ionic and electronic transport in apatite-type $La_{10-x}(Si,Al)_6O_{26\pm\delta}$. *J. Solid State Chem.* **178,** 2050–2061 (2005).

76. Leon-Reina, L. *et al*. High oxide ion conductivity in Al-doped germanium oxyapatite. *Chem. Mater.* **17,** 596–600 (2005).

77. Masabuchi, U., Higuchi, M., Takeda, T. & Kikkawa, S. Oxide ion conduction mechanism in RE$_{9.33}$(SiO$_4$)$_6$O$_2$ and Sr$_2$RE$_8$(SiO$_4$)$_6$O$_2$ (RE = La, Nd) from neutron powder diffraction. *Solid State Ion.* **177,** 263–268 (2006).

78. Yoshioka, H. Enhancement of ionic conductivity of apatite-type lanthanum silicates doped with cations. *J. Am. Ceram. Soc.* **90,** 3099–3105 (2007).

79. An, T. *et al*. Crystallographic correlations with anisotropic oxide ion conduction in aluminum-doped neodymium silicate apatite electrolytes. *Chem. Mater.* **25,** 1109–1120 (2013).

80. Arikawa, H., Nishigchi, H., Ishihara, T. & Takita Y. Oxide ion conductivity in Sr-doped La$_{10}$Ge$_6$O$_{27}$ apatite oxide. *Solid State Ion.* **136–137,** 31–37 (2000).

81. Wei, T., Xu, J. & Zhu, W. New apatite-type oxide ion conductors Ce$_{9.33+x}$Si$_6$O$_{26+\delta}$: Structures, phase stabilities, electrical properties, and conducting mechanisms. *Energy Sci. Eng.* **10,** 525–537 (2022).

82. Martin-Sedeño, M. C. *et al*. Enhancement of Oxide Ion Conductivity in Cuspidine-Type Materials. *Chem. Mater.* **16,** 4960–4968 (2004).

83. Chesnaud, A. *et al*. Cuspidine-like compounds Ln$_4$[Ga$_{2(1-x)}$Ge$_{2x}$O$_{7+x}\square_{1-x}$]O$_2$ (Ln = La, Nd, Gd; $x \leq 0.4$). *Chem. Mater.* **16,** 5372–5379 (2004).

84. Martin-Sedeño, M. C. *et al*. Structural and electrical investigation of oxide ion and proton conducting titanium cuspidines. *Chem. Mater.* **17,** 5989–5998 (2005).

85. Thomas, C. I. *et al*. Phase stability control of interstitial oxide ion conductivity in the La$_{1+x}$Sr$_{1-x}$Ga$_3$O$_{7+x/2}$ melilite family. *Chem. Mater.* **22,** 2510–2516 (2010).

86. Diaz-Lopez, M. *et al*. Interstitial Oxide Ion Conductivity in the Langasite Structure: Carrier Trapping by Formation of (Ga,Ge)$_2$O$_8$ Units in La$_3$Ga$_{5-x}$Ge$_{1+x}$O$_{14+x/2}$ ($0 < x \leq 1.5$). *Chem. Mater.* **15,** 5742–5758 (2019).

87. Li, X. *et al*. B-site mixed cationic tetrahedral layer confined the concentration and mobility of interstitial oxygen in mellite family. *J. Mater. Chem. A* **11,** 5615–5626 (2023).

88. Bazzaoui, H. *et al*. La Substitution into the Melilite Derivative $Ca_5Ga_6O_{14}$: Prediction, Synthesis and Ionic Conductivity. *Inorg. Chem.* **63,** 18902–18913 (2024).

89. Rotman, S. R. & Tuller, H. L. Defect-property correlations in garnet crystals. *VII: The electrical conductivity and defect structure of yttrium aluminum and yttrium iron garnet solid solutions. *J. Electroceram.* **2,** 95–104 (1998).

90. Kharton, V. V. *et al*., Ionic transport in $Gd_3Fe_5O_{12}$- and $Y_3Fe_5O_{12}$-based garnets. *J. Electochem. Soc.* **150,** J33–J42 (2003).

91. Kramer, S. A. & Tuller, H. L. A novel titanate-based oxygen ion conductor: $Gd_2Ti_2O_7$. *Solid State Ion.* **82,** 25–23 (1995).

92. Morkhova, Y. A. *et al*. Magnocolumbites $Mg_{1-x}M_xNb_2O_{6-\delta}$ ($x$ = 0, 0.1, and 0.2; $M$ = Li and Cu) as new oxygen ion conductors: Theoretical assessment and experiment. *J. Phys. Chem. C* **127,** 52–58 (2023).

93. Singh, P. & Goodenough, J. B. $Sr_{1-x}K_xSi_{1-y}Ge_yO_{3-0.5x}$: A new family of superior oxide-ion conductors. *Energy Environ. Sci.* **5,** 9626–9631 (2012).

94. Singh, P. & Goodenough, J. B. Monoclinic $Sr_{1-x}Na_xSiO_{3-0.5x}$: New superior oxide ion electrolytes. *J. Am. Chem. Soc.* **135,** 10149–10154 (2013).

95. Wei, T. *et al*. $Sr_{3-3x}Na_{3x}Si_3O_{9-1.5x}$ ($x$ = 0.45) as a superior solid oxide-ion electrolyte for intermediate temperature-solid oxide fuel cells. *Energy Sci. Eng.* **7,** 1680–1684 (2014)

96. Tealdi, C. *et al*. Nature of conductivity in $SrSiO_3$-based fast ion conductors. *Chem. Commun.* **50,** 14732–14735 (2014).

97. Unke, O. T. *et al*. Machine learning force fields. *Chem. Rev.* **121,** 10142–10186 (2021).

98. Wang, G. et al. Machine learning interatomic potential: Bridge the cap between small-scale models and realistic device-scale simulations. *iScience* **27,** 109673 (2024).

99. Xia, J., Zhang, Y. & Jiang, B. The evolution of machine learning potentials for molecules, reactions and materials. *Chem. Soc. Rev.* **54,** 4790–4821 (2025).

100. *Web of Science* https://www.webofscience.com/wos/woscc/smart-search (2025).

101. *Scopus* https://www.elsevier.com/products/scopus (2025).

102. *Google Scholar* https://scholar.google.com/ (2025).

103. Shannon, R. D. Revised effective ionic radii and systematic studies of interatomic distances in halides and chalcogenides. *Acta Crystallogr. A* **32,** 751–767 (1976).

104. Baloch, A. A. B. *et al*. Extending Shannon's ionic radii database using machine learning. *Phys. Rev. Materials* **5,** 043804 (2021).

105. Jang, S.-H. *GoodRegressor*: A general-purpose symbolic regression framework for physically interpretable materials modeling. Preprint at https://doi.org/10.48550/arXiv.2510.18325 (2025).

# Supplementary Information

# Charting the Landscape of Oxygen Ion Conductors: A 60-Year Dataset with Interpretable Regression Models

*Seong-Hoon Jang*[*1] *, Shin Kiyohara*[1]*, Hitoshi Takamura*[2]*, Yu Kumagai*[**1,3]

[1] *Institute for Materials Research*, Tohoku University, 2-1-1 Katahira, Aoba-ku, Sendai 980-8577, Japan.

[2] Department of Materials Science, Graduate School of Engineering, Tohoku University, 6-6-02 Aramaki, Aoba-ku, Sendai 980-8579, Japan

[3] Organization for Advanced Studies, Tohoku University, 2-1-1 Katahira, Aoba-ku, Sendai 980-8577, Japan

*Corresponding authors:

jang.seonghoon.b4@tohoku.ac.jp (S.-H. Jang)

yukumagai@tohoku.ac.jp (Y. Kumagai)

This file contains:

- Correction of the Nernst-Einstein Relation in Arrhenius Plot Analysis

**Correction of the Nernst-Einstein Relation in Arrhenius Plot Analysis**

A common issue in the literature involves the use of an incorrect form of the Nernst–Einstein equation when constructing Arrhenius plots. The correct relationship is given by

$$\sigma_T T = A \exp\left(-\frac{E_a}{RT}\right), \qquad (S1)$$

where $R$ is the gas constant. However, many studies instead employ the simplified form

$$\sigma_T = A \exp\left(-\frac{E_a}{RT}\right). \qquad (S2)$$

This inconsistency leads to systematic errors in the calculated values of $E_a$ and $A$. For instance, in an experimental study on $Ce_{9.33+x}Si_6O_{26+\delta}$, the incorrect expression yielded $E_a = 0.95$, 0.97, and 0.96 eV for $x = 0.43$, 0.67, and 1.03, respectively (see Fig. 7 in that paper).[1] Upon manually re-curating the Arrhenius data using the correct form, the corresponding activation energies were found to be $E_a = 0.92$, 0.93, and 0.92 eV. Thus, the incorrect formulation systematically overestimates $E_a$ by approximately 0.03~0.04 eV. Although such deviations may appear minor relative to the magnitude of $E_a$, they can nonetheless compromise the accuracy of datasets and hinder reliable model development. Therefore, careful attention to the correct equation form is warranted.

**A Brief Summary of Symbolic Regression Tool *GoodRegressor***

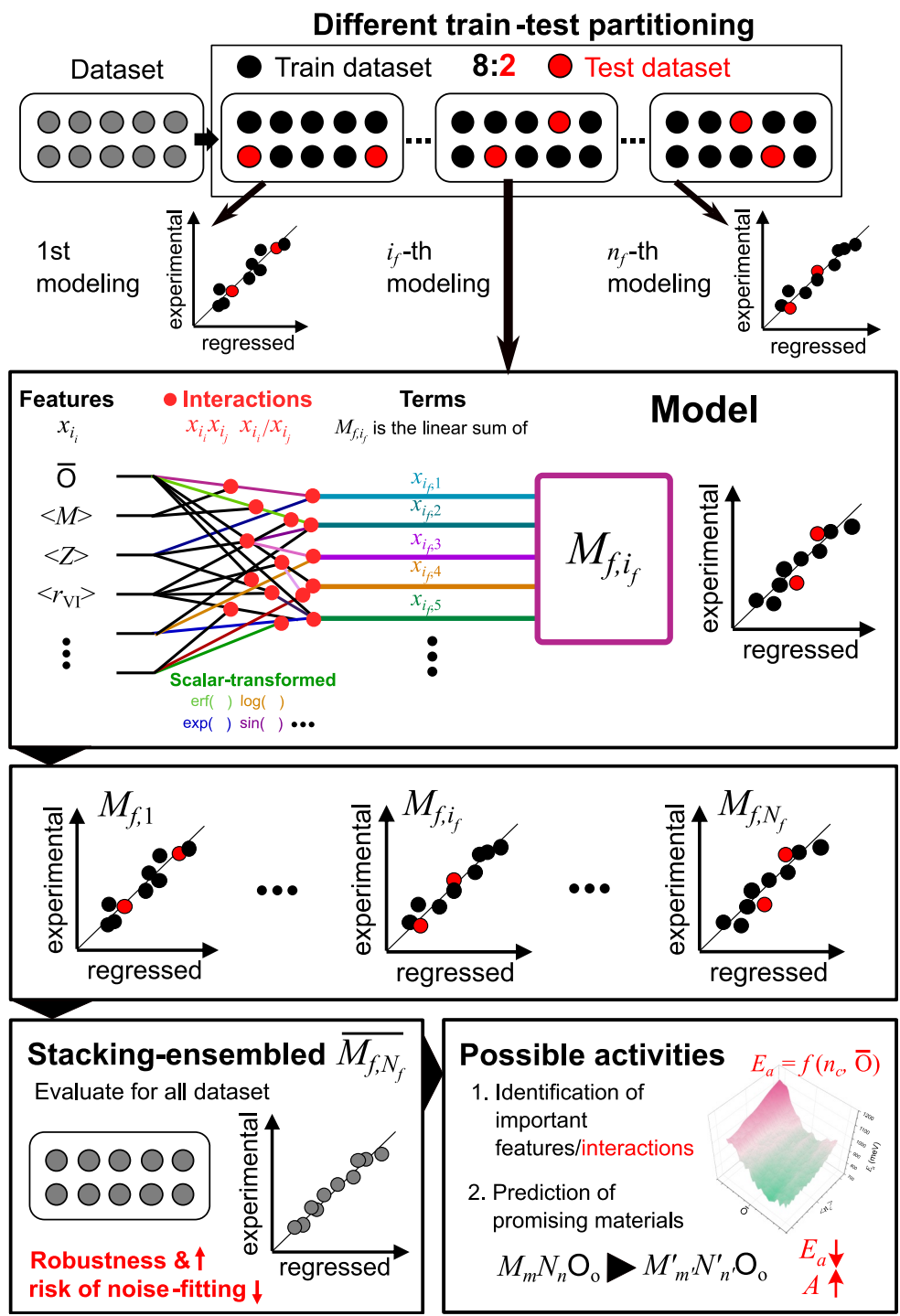

**Supplementary Fig. 1** Workflow of the symbolic regression used in this study. Starting from the experimental dataset, the program performs multiple train-test partitions (8: 2 ratio, repeated ten

times) to construct stacking-ensembled symbolic regression models.[3] In each iteration, feature interactions (e.g., multiplications $x_{i_i}x_{i_j}$ and divisions $x_{i_i}/x_{i_j}$) and scalar transformations (e.g., $\text{erf}(x_{i_i})$, $\log(x_{i_i})$, $\exp(x_{i_i})$, $\sin(x_{i_i})$, $\sqrt{x_{i_i}}$, $x_{i_i}^2$, $\cdots$) are explored to maximize the coefficient of determination ($R^2_{\text{test}}$) of the test dataset. The resulting models are stacking-ensembled to yield a robust and interpretable expression that mitigates overfitting. By analyzing the frequency of an interaction and the magnitude of its $z$-scored coefficients across the ensembles, the program identifies key physical descriptors governing the target property and enables the virtual design of promising material candidates through compositional modulation.

In *GoodRegressor* as illustrated in **Supplementary Fig. 1**, starting from the experimental dataset, the program performs random train-test partitions with a user-defined ratio (8: 2 in this study) to build symbolic regression models.[2] This process is repeated $n_f$ times (10 times in this study) using different randomly sampled training datasets of the same size drawn from the full dataset, thereby enabling "bagging". In the $i_f$-th modeling iteration ($1 \leq i_f \leq n_f$), beginning with the feature set $x_{i_i}$ listed in **Table 2 in the main text**, the program optimizes not only their multiple interactions (e.g., multiplications $x_{i_i}x_{i_j}$ and divisions $x_{i_i}/x_{i_j}$) but also scalar transformations (e.g., $\text{erf}(x_{i_i})$, $\log(x_{i_i})$, $\exp(x_{i_i})$, $\sin(x_{i_i})$, $\sqrt{x_{i_i}}$, $x_{i_i}^2$, $\cdots$) to maximize the coefficient of determination ($R^2_{\text{test}}$) of the test dataset based on the regression model trained on the training subset. The resulting symbolic regression model $M_{f,i_f}$ is expressed as a linear combination of terms ($x_{i_f,1}, x_{i_f,2}, \cdots,$), each corresponding to a feature, an interaction, or their scalar-transforms (often complicated algebraically). Finally, all iteration models ($M_{1,i_f}, \cdots, M_{f,i_f}, \cdots, M_{f,N_f}$) are stacking-ensembled to yield the final model $\overline{M_{f,n_f}}$. Increasing $n_f$ enhances model robustness, reflected in higher $R^2$ values for predictions across the entire dataset, and mitigates the risk of noise-fitting that could otherwise lead to overfitting on unseen data. On top of this, from the ensemble of models $\{M_{1,i_f}, \cdots, M_{f,i_f}, \cdots, M_{f,n_f}\}$, both important individual features and key "interactions" can be identified by evaluating their frequency of occurrence and average coefficient magnitudes across models. Roughly, the former answers to "how many models do have term(s) where target features, $x_{i_i}$, $x_{i_j}$, $\cdots$, coexist (interact)?", while the latter answers to "what is the total average of the absolute values of $z$-scored coefficients for term(s) where target features, $x_{i_i}$, $x_{i_j}$, $\cdots$, coexist (interact)?". Moreover, the stacking-ensembled $\overline{M_{f,n_f}}$ can be used to predict promising material candidates with desired target properties by virtually substituting or doping atoms (ions) within existing compositions in the experimental dataset.

**Supplementary Reference**

1. Wei, T., Xu, J. & Zhu, W. New apatite-type oxide ion conductors $Ce_{9.33+x}Si_6O_{26+\delta}$: Structures, phase stabilities, electrical properties, and conducting mechanisms. *Energy Sci. Eng.* **10,** 525–537 (2022).

2. Jang, S.-H. *GoodRegressor*: A general-purpose symbolic regression framework for physically interpretable materials modeling. Preprint at https://doi.org/10.48550/arXiv.2510.18325 (2025).